\documentclass[letterpaper]{article}
\usepackage{spconf,amsmath,amssymb,graphicx,array,booktabs, etoolbox}
\usepackage[hidelinks]{hyperref}

\newcommand{\Hvs}{H_{\mathrm{v\text{-}sp}}}
\newcommand{\Hcs}{H_{\mathrm{c\text{-}sp}}}

\begin{document}
\ninept
\title{PLACE: Positional Latent Adaptation via Conditioned
Embeddings for Binaural Audio Generation}
\name{Tiernon Riesenmy$^{\star \dagger}$\thanks{Work completed while T. Riesenmy was an intern at Dolby Laboratories.} \qquad You Zhang$^{\dagger}$ \qquad
Gautam Bhattacharya$^{\dagger}$ \qquad Andrea Fanelli$^{\dagger}$}
\address{\small
$^{\star}$The Wallace H. Coulter Department of Biomedical Engineering,
\small Georgia Institute of Technology, Atlanta, GA, USA\\
\small $^{\dagger}$Advanced Technology Group, Dolby Laboratories, San Francisco, CA, USA}
\maketitle

\begin{abstract}
We present PLACE, a method that extends the pretrained any-to-audio model AudioX for binaural generation from arbitrary combinations of text, video, and optional audio prompts. PLACE augments video conditioning with Perception Encoder Core features, aligns text and video representations to derive spatial cues, and applies a conditioning-dependent low-rank transformation to the generated latent. The adapter is supervised via decoded-audio interaural level and time difference objectives. Trained on MRSAudio, PLACE improves most metrics over ViSAGe on FAIR-Play and achieves an improved SpatialCLAP score over SpatialSonic on the BEWO-1M Single Static test split. Listener evaluations favor PLACE for video-to-audio and out-of-distribution text-to-audio generation, demonstrating flexible multimodal control and improved spatial consistency.

\end{abstract}

\section{Introduction}
Spatial audio enhances the level of immersion \cite{10.3389/frsip.2022.904866, garcia2026emorsion} in virtual reality (VR), augmented reality (AR), and cinematic experiences.  Binaural audio remains one of the most accessible formats of spatial audio since playback of these recordings only requires a pair of headphones. Such recordings encode directional cues for human hearing, most notably interaural level differences (ILD) and interaural time differences (ITD) between the left and right channels. These experiences increasingly involve accompanying video and text, so a binaural audio generation model that conditions on this surrounding multimodal context is essential for producing spatial audio that matches the scene.

Recent work has successfully generated spatial audio in various formats, including binaural audio, from input prompts. ViSAGe~\cite{kim2025visage} generates first-order ambisonics (FOA) from video and camera direction. This method, while it performs well, does not support flexible, multimodal controls over spatial audio generation. Similarly, FoleySpace~\cite{zhao2026foleyspace} generates binaural audio using estimated 3D source trajectories and monaural audio conditioning but does not support text inputs. SpatialSonic \cite{sun2024both} generates spatial stereo audio given an image or text that provides spatial context, but relies on external LLM queries at inference for optimal performance. Other works in the field have made strides towards a unified any-to-spatial audio model but either do not support spatially descriptive text \cite{zhang2025visaudioendtoendvideodrivenbinaural} or focus predominantly on the video input with an optional spatial text conditioning path~\cite{lei2026swansphere}.  

A distinct lack of any-to-spatial audio models primarily stems from the limited availability of data that pairs spatial audio, video, and spatially descriptive text. Training such a model from scratch would require a substantial training corpus beyond what is currently available. One could generate a synthetic dataset, but such data cannot accurately reflect real-world contexts. Creating a suitable dataset would require either careful human annotation or human refinement of LLM-generated annotations, both of which are costly. We propose a method that leverages existing architectures for transfer learning using a properly annotated dataset containing aligned binaural audio, field-of-view video, and spatially-annotated text.

In this work, we introduce \textbf{P}ositional \textbf{L}atent \textbf{A}daptation via \textbf{C}onditioned \textbf{E}mbeddings (\textbf{PLACE}). Unlike prior work, our method supports any combination of video, text, and audio for binaural audio generation. \hspace{-2pt}Our method places sound by adapting the latent of a pretrained generator rather than through an intermediate inference stage ~\cite{zhao2026foleyspace, sun2024both}. We evaluate our model's generalization on three test splits from FAIR-play \cite{gao2019visualsound} and the BEWO-1M Single Static (BEWO-1M SS-set) test split \cite{sun2024both}, and conduct a subjective listener test to evaluate the perceptual quality of the generated audio based on an input prompt. We achieve competitive results across all evaluations, demonstrating that our data-efficient methodology successfully generalizes across multiple spatial audio generation tasks.

\section{Proposed method: PLACE}

\begin{figure*}[t]
\centering
\includegraphics[width=0.86\textwidth]{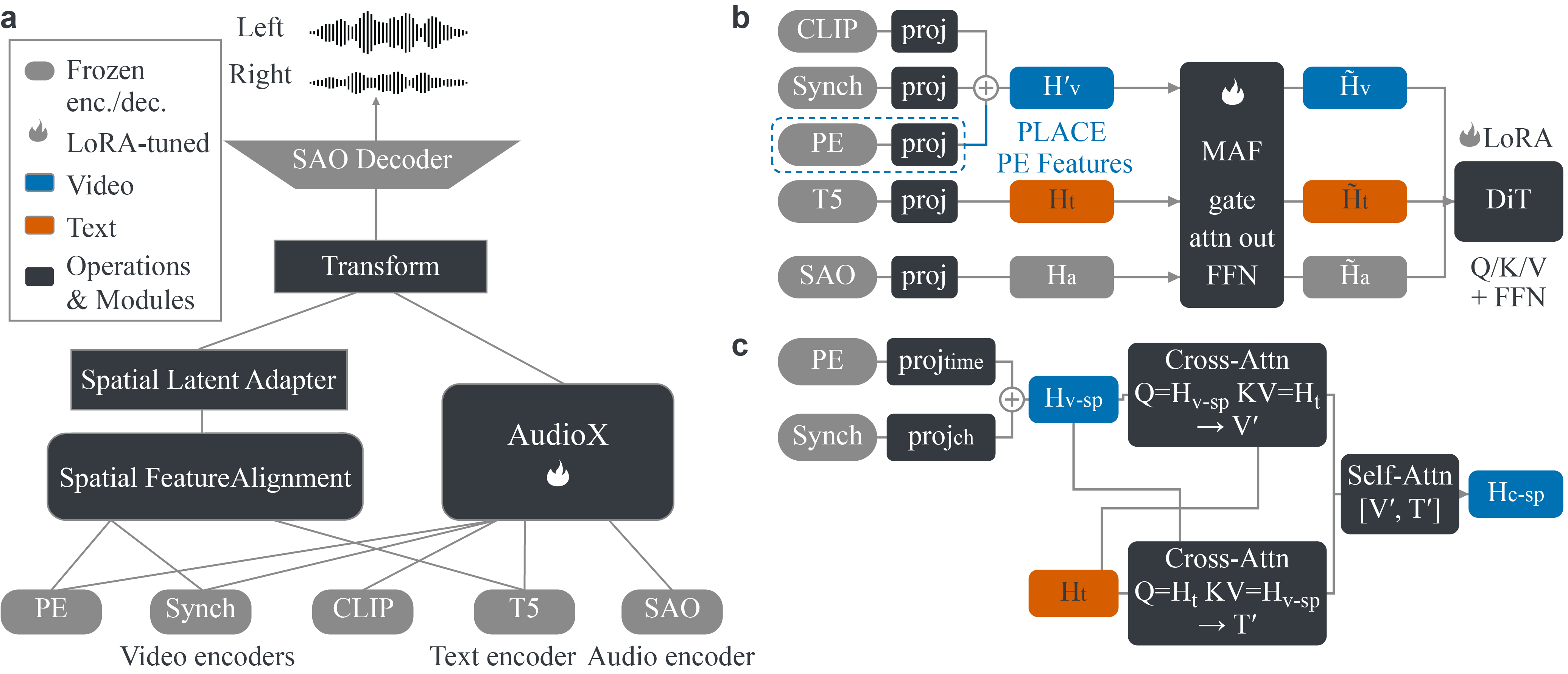}
\caption{Overview of our proposed PLACE framework. We emphasize the video feature path in blue and the text feature path in orange. \textbf{a.} AudioX generates a latent; spatial conditioning drives its transformation before the frozen SAO decoder.
All five pretrained encoders and the decoder remain frozen; AudioX fine-tuned via LoRA. \textbf{b.} Projected PE Core features augment
CLIP and Synchformer features before the MAF module. \textbf{c.} Text--video cross-attention and joint self-attention yield spatial features $\Hcs$.}
\label{fig:framework_overview}
\end{figure*}

% \NZ{Will get back here later}
As shown in Fig.~\ref{fig:framework_overview}, PLACE extends AudioX~\cite{tian2025audiox} for binaural audio generation through four contributions: augmenting video conditioning with PE-Core features (Sec.~\ref{sec:pe}), aligning text and video in a spatial branch (Sec.~\ref{sec:biattn}), adapting the latent estimate with a low-rank transform (Sec.~\ref{sec:sla}), and training the spatial modules with interaural level and time difference losses (Sec.~\ref{sec:losses}).
% Together, these contributions extend the original AudioX architecture to binaural audio generation conditioned on multimodal context.
% PLACE extends AudioX for binaural audio generation through four contributions that operate in sequence, illustrated in Fig.~\ref{fig:framework_overview}. First, we augment AudioX's video conditioning with PE-Core features (Sec.~\ref{sec:pe}). In parallel, we introduce a dedicated spatial branch, where PE-Core and Synchformer video features are cross-attended against T5 text features to produce spatial conditioning $H_{c\text{-sp}}$ (Sec.~\ref{sec:biattn}). This spatial conditioning drives a low-rank adapter that transforms the clean-latent estimate recovered from the fine-tuned AudioX diffusion transformer into a spatially adapted latent (Sec.~\ref{sec:sla}), which is then convereted to spatial audio by the frozen SAO decoder. Finally, we train the alignment block and adapter using interaural level and time difference losses computed directly on this decoded audio (Sec.~\ref{sec:losses}), so spatial supervision reaches the model through the same audio a listener would hear.
% Together, these contributions extend the original AudioX architecture to binaural audio generation conditioned on multimodal context.

\subsection{Preliminary Study on AudioX Backbone}
\label{sec:prelim}

AudioX is an any-to-audio framework that generates audio conditioned on any combination of text, video, and audio prompts. It operates on the latent space of a pre-trained Stable Audio Open (SAO)~\cite{evans2025stable} autoencoder, denoising an SAO audio latent with a diffusion transformer (DiT). The conditioning comes from a multimodal adaptive fusion (MAF) module, which combines Synchformer~\cite{synchformer2024iashin} and frame-wise CLIP-ViT-B/32~\cite{radford2021learning} video features, T5~\cite{T5} text features, and SAO audio latents into a single representation $H_c$ that drive the DiT for audio generation. 

\textbf{Why AudioX}. SAO's autoencoder operates on stereo audio, giving it the correct shape for our goal of generating binaural audio. We hypothesize that, though trained on stereo audio, the SAO latents preserve binaural spatial cues to some extent, so that we may adapt
and strengthen this information and align it with multimodal inputs. To validate such spatial preservation, we encode and decode two binaural recordings through the SAO autoencoder, and measure differences in level, timing, and similarity between the left and right channels across 32 ERB-spaced frequency bands. Despite imperfect waveform reconstruction, the decoded audio largely retains its spatial width (side-channel energy changes by at most 0.22 dB) and closely follows the original left--right level differences ($\text{Pearson coeff.} = 0.95$--$0.96$). Informal listening also suggests that the decoded audio retains spatial cues consistent with the input. 
These observations support using SAO latent as a base representation for binaural audio generation. 
% can encode binaural characteristics, motivating the work outlined in (Sec.~\ref{sec:sla}). 
Together with AudioX's multimodal conditioning, we believe AudioX provides a solid base architecture well-suited for our goal.

\subsection{\hspace{-1pt}Video Feature Augmentation with Perception Encoder Core}
\label{sec:pe}
Synchformer and frame-by-frame CLIP-ViT-B/32 in the video path of original AudioX are both image-based encoders. They capture object composition within individual frames but lack an explicit representation of spatial placement among objects, or of how sound sources move across frames, both of which matter for binaural generation.
We therefore augment the video path with PE-Core-G14-448 from Meta's Perception Encoder (PE) family~\cite{bolya2025PerceptionEncoder}. Its features support downstream spatial tasks and achieve an MSR-VTT text-to-video retrieval score
of 51.2~\cite[Table~4]{bolya2025PerceptionEncoder}. For context, a published evaluation reports a score of 30.4 for B-scale CLIP~\cite[Table~9]{wang2024internvideo2}. We select PE-Core over PE-Spatial as the latter's $32\times32$ patch grid requires \(1024\times\) the storage of a single global descriptor at the same feature dimensionality and precision, increasing our training-set cache from gigabytes to terabytes. We learn binaural conditioning from PE-Core via spatial feature alignment (Sec.~\ref{sec:biattn}) and interaural objectives (Sec.~\ref{sec:losses}), as illustrated in Fig.~\ref{fig:framework_overview}b.

\subsection{Spatial Feature Alignment}
\label{sec:biattn}
Since AudioX did not enforce binaural audio generation, the generated latent does not explicitly encode spatial cues, though it preserves some as discussed in Sec.~\ref{sec:prelim}. Retraining the DiT is infeasible without a large corpus, so we inject spatial cues from the input prompts into the generated latent via a learned transform. We maintain T5 text features,
supported by a prior finding that T5 outperforms CLAP~\cite{laionclap2023}
for spatial audio generation~\cite{sun2024both}. We add two projection layers, one after Synchformer and the other to temporally resample PE Core features; the video path becomes:
$\Hvs=\operatorname{proj}_{\mathrm{time}}(H_{\mathrm{PE}})+\operatorname{proj}_{\mathrm{ch}}(H_{\mathrm{Synch}})$,
where $H_{\mathrm{Synch}}\in\mathbb R^{240\times768}$,
$H_{\mathrm{PE}}\in\mathbb R^{50\times1280}$. The projections map channels and time, providing $\Hvs\in\mathbb R^{240\times1280}$. Three four-head attention blocks align the modalities (Fig.~\ref{fig:framework_overview}c).
Padded text is masked when serving as keys or values and projected to width 1280.
Retaining the first $T_{\mathrm{sync}}$ joint-output positions yields $\Hcs\in\mathbb R^{240\times1280}$.
For text-only prompts, learned projection biases and a video-modality embedding provide a fixed \(T_{\mathrm{sync}}\times1280\) no-video initialization; subsequent attention makes the spatial representation text-dependent.

\subsection{Spatial Latent Adapter}
\label{sec:sla}
PLACE learns the transform derived from spatial cues inherent to the input prompts via an MLP after the spatial feature alignment (Fig~\ref{fig:framework_overview}a). The MLP maps $\Hcs$, linearly resampled from 240 frames to latent length,
to position-dependent low-rank factors and a channel-wise gate:
\begin{align*}
(A_\ell,B_\ell)&=\operatorname{MLP}_{AB}(\Hcs{}_{,\ell}),\quad
 g_\ell=\operatorname{MLP}_{g}(\Hcs{}_{,\ell}),\\
z'_{0,\ell}&=z_{0,\ell}+g_\ell\odot
 [\operatorname{GELU}(z_{0,\ell}A_\ell)B_\ell].
\end{align*}
Here $A_\ell\in\mathbb R^{d\times r}$,
$B_\ell\in\mathbb R^{r\times d}$, $g_\ell\in\mathbb R^d$, and $\odot$
is channel-wise multiplication; GELU is the Gaussian error linear unit. \(\ell\) indexes the \(T\) latent-time positions after temporally resampling. We use a bottleneck rank of $r=16$. With this bottleneck, the two factors require $2 \times d \times r$ predicted coefficients per latent position, compared with $d^2$ for a full channel transformation.
At inference, the adapter acts once on the final denoised latent before
decoding. 

During training, the adapter acts on a single-step clean-latent estimate
from a random timestep, avoiding a full sampling trajectory. With
$z_t=\alpha_tz_0+\sigma_t\epsilon$ and
$(\alpha_t,\sigma_t)=(\cos(\pi t/2),\sin(\pi t/2))$, the velocity target is
$v_t=\alpha_t\epsilon-\sigma_tz_0$. We train the adapter with
% \begin{align*}
\begin{equation}
\mathcal L_{\mathrm{lat}}=\mathbb E_{z_0,\epsilon,t}
 \|v_t-v_\theta(z_t,t,H_c)\|_2^2 ,
\label{eq:llat}
\end{equation}
and recover $\hat z_0$ with $\hat{z}_0=\alpha_tz_t-\sigma_tv_\theta(z_t,t,H_c).$ The adapter transforms $\hat z_0$. The $\alpha_i^2$ weights in the losses from Sec.~\ref{sec:losses} down-weight high-noise steps,
at which $\hat{x}_0$ poorly approximates the latent that is decoded at inference.

\subsection{Interaural Losses and Training}
\label{sec:losses}
To further guide PLACE towards the generation of binaural audio, we
introduce two losses that directly supervise the interaural cues
responsible for spatial perception, following work in target sound extraction~\cite{losses} and neural
acoustic fields~\cite{ick2024spatiallyaware}. We compute both losses on decoded audio for a subset of each batch for efficiency, which provides the spatial feature alignment block and latent adapter along the decoding path with a training signal grounded in the audio a listener would actually hear.

The first loss supervises interaural level differences (ILD), which convey the relative loudness of a sound source between the two ears:
\begin{equation}
\mathcal L_{\mathrm{ILD}}=\frac1{N_{\mathrm{sub}}T}
 \sum_{i=1}^{N_{\mathrm{sub}}}\sum_{t=1}^T\alpha_i^2
 (D_{i,t}-D^{\mathrm{GT}}_{i,t})^2,
\label{eq:ild_loss}
\end{equation}
where $D_{i,t}=\log_{10}[(P_{L,i,t}+\varepsilon)/(P_{R,i,t}+\varepsilon)]$,
$P_L,P_R$ are framewise mean channel powers, $\varepsilon=10^{-7}$, and
$D^{\mathrm{GT}}$ is the ground-truth log-power ratio. 

The second loss supervises interaural time differences (ITD), which convey directional cues through the arrival-time offset between the two ears:
\begin{equation}
\mathcal L_{\mathrm{ITD}}=\frac{\displaystyle
 \sum_{i=1}^{N_{\mathrm{sub}}}\sum_{t=1}^{T}\sum_{k=-\tau_{\max}}^{\tau_{\max}}
 \alpha_i^2(R_{i,t}[k]-R^{\mathrm{GT}}_{i,t}[k])^2}
 {N_{\mathrm{sub}}T(2\tau_{\max}+1)},
\label{eq:itd_loss}
\end{equation}
where $R,R^{\mathrm{GT}}$ are predicted and target generalized
cross-correlation with phase transform (GCC-PHAT)~\cite{knapp1976generalized}
coefficients at lag $k$. In both losses, $N_{\mathrm{sub}}$ the sub-batch size, $T$ is the
number of audio frames, and $\alpha_i$ is the sample's noise-schedule
coefficient. Both signals use native 44.1 kHz decoder output, 512-sample
windows (11.6 ms), 256-sample hops (5.8 ms), and $\tau_{\max}=13$
($\pm0.29$ ms). Identical framing makes the objective well-posed, but it
supervises only the central part of the physiological $\pm0.8$ ms range.
The evaluation ITD metric instead covers the full range at 16 kHz. 

We combine these losses with the latent-matching objective
$\mathcal{L}_{\mathrm{lat}}$ (Eq.~\ref{eq:llat}) into a single total
objective:
\begin{equation}
\mathcal{L} = \mathcal{L}_{\mathrm{lat}} +
 \lambda_{\mathrm{ILD}}\mathcal{L}_{\mathrm{ILD}} +
 \lambda_{\mathrm{ITD}}\mathcal{L}_{\mathrm{ITD}}.
\label{eq:total_loss}
\end{equation}

All five pretrained encoders and the SAO decoder remain frozen
throughout training. Because the spatial alignment block and adapter lie
on the decode path and receive gradients only through
$\mathcal{L}_{\mathrm{ILD}}$ and $\mathcal{L}_{\mathrm{ITD}}$, we train
PLACE in two stages. We first fine-tune AudioX for 10 epochs under
$\mathcal{L}_{\mathrm{lat}}$ alone, updating rank-16
LoRA~\cite{hu2022lora} adapters on the linear and attention layers and
the gate network together with the video conditioning projections. We
then attach the spatial modules and train for 10 further epochs under
the total objective $\mathcal{L}$ in Eq.~\ref{eq:total_loss}, with
$\lambda_{\mathrm{ILD}}=\lambda_{\mathrm{ITD}}=1$, $N_{\mathrm{sub}}=6$,
batch size 24, and learning rate $3\!\times\!10^{-5}$.

% \newpage
\section{Experiments and Results}
% We evaluate our model's generalization ability on out-of-distribution (OOD) datasets to determine whether our data-efficient methodology enables the pretrained architectures to generalize effectively to binaural audio generation. We then conduct ablation studies on a held-out dataset from our training corpus to quantify the contribution of each component. Finally, we assess the perceptual quality of the generated audio through subjective listening tests.
A central claim of our approach is that it is data efficient: by adapting a pretrained any-to-audio model rather than training from scratch, PLACE achieves strong spatial audio generation despite fine-tuning on a comparatively small annotated corpus. We evaluate this claim on datasets entirely outside our training distribution (Sec.~\ref{sec:ood}), then decompose which components drive in-distribution performance through ablations (Sec.~\ref{sec:ablat}), and assess perceptual quality through listener studies (Sec.~\ref{sec:test}).

\subsection{Experimental Setup}
\textbf{Datasets}. MRSAudio \cite{guo2025mrsaudio} is an aligned multimodal dataset that contains paired spatially-descriptive text, field-of-view video, and binaural audio. 
We use the MRSSound subset and augment it from 12,057 to 24,114 annotated 10s clips by jointly mirroring videos, swapping audio channels, and exchanging directional caption terms. 596 disjoint clips are held out for ablations. On OOD tasks, we use the first three test splits from FAIR-Play dataset for video-to-spatial audio (V2A) generation and BEWO-1M SS-set test split for text-to-spatial audio (T2A) generation. The FAIR-Play test splits each contain 187 binaural recordings of musicians in a music room paired with field-of-view videos. BEWO-1M SS-set test split consists of 4301 text-spatial stereo audio pairs.

\textbf{Baselines}. AudioX is a natural baseline. ViSAGe serves as our V2A baseline for comparison as it is a pioneering work for spatial audio generation and there is a lack of publicly available binaural audio generation models. It was pretrained on VGGSound for mono audio generation, and fine-tuned on their YT-Ambigen dataset consisting of 81,594 5-second clips. Since ViSAGe generates FOA, we render its output binaurally using \texttt{spaudiopy} v0.2.0~\cite{hold2025spaudiopy}, applying \texttt{sofa\_to\_sh} at $N_{\mathrm{sph}} = 1$ followed by \texttt{sh2bin}. We use the NH12 listener HRTFs from the ARI HRTF database~\cite{arihrtf}, distributed using the \texttt{SimpleFreeFieldHRIR} 1.0 convention of SOFA/AES69~\cite{aes69}.
SpatialSonic exhibits strong text-to-audio (T2A) performance and serves as our baseline for the task. The model uses the same pretrained text encoder and audio autoencoder as our method, and was trained on over 1M audio-text pairs.

\textbf{Metrics}.
We evaluate with four metrics, for all of which lower is better. Fr\'echet Stereo
Audio Distance (FSAD) \cite{sun2024both} is a spatially-informed variant of the
Fr\'echet Audio Distance (FAD) \cite{cheng2025taming,
tian2025audiox, kim2025visage}; we train our own StereoCRW \cite{chen2022sound}
checkpoint on MRSAudio to compute it. We additionally report per-clip interaural cue
distribution distances: for both target and generated audio we extract ITD via
GCC-PHAT and ILD as in (eq.~\ref{eq:ild_loss}), then take the Wasserstein distance
between the resulting ILD and ITD distributions. Finally, SpatialCLAP \cite{11460767}
supplies audio and text encoders trained with spatial contrastive learning; we report
the difference between the target and generated audio similarity scores against the
provided caption.

\begin{table}[t]
\centering
\small
\caption{Out-of-distribution objective evaluation on FAIR-Play (video-to-audio,
three test splits) and the BEWO-1M SS-set test split (text-to-audio). Lower is
better for all metrics; ITD and ILD are reported in ms and dB; SCLAP is the
SpatialCLAP difference from ground truth. Bold marks the better score between PLACE and the baseline
in each row. \vspace{2pt}}
\label{tab:ood}
\setlength{\tabcolsep}{3pt}
\begin{tabular*}{\linewidth}{@{\extracolsep{\fill}}lrrrr@{}}
\toprule
\multicolumn{5}{@{}l}{\textit{FAIR-Play V2A}}\\
Split / Model & FSAD & ITD & ILD & SCLAP\\
\midrule
1 / ViSAGe & 15.9016 & 0.1136 & 2.1306 & --\\
1 / {PLACE} & \textbf{15.2879} & \textbf{0.1077} & \textbf{2.0495} & --\\
2 / ViSAGe & 14.8996 & \textbf{0.1133} & 2.0198 & --\\
2 / {PLACE} & \textbf{14.3745} & 0.1217 & \textbf{1.9827} & --\\
3 / ViSAGe & 14.2590 & 0.1228 & 2.1849 & --\\
3 / {PLACE} & \textbf{13.8444} & \textbf{0.1116} & \textbf{2.1395} & --\\
\midrule
\midrule
\multicolumn{5}{@{}l}{\textit{BEWO-1M SS-set test split T2A}}\\
Model & FSAD & ITD & ILD & SCLAP\\
\midrule
SpatialSonic & \textbf{6.4710} & \textbf{0.2494} & \textbf{0.7843} & -0.001\\
{PLACE} & 7.0523 & 0.3182 & 1.5638 & \textbf{-0.050}\\
\bottomrule
\end{tabular*}
\end{table}

\subsection{Generalization Under Limited Training Data}
\label{sec:ood}
Results for our OOD evaluations are provided in Table~\ref{tab:ood}. On V2A we
improve on ViSAGe for every metric on each split except ITD distribution distance on
Split 2. On T2A we are competitive on all metrics but ILD distribution distance and
even outperform SpatialSonic on the SpatialCLAP score, despite SpatialSonic having been trained on a
larger corpus whose distribution matches the test set. For comparison, PLACE trains on roughly 3.4x less data than ViSAGe and up to 41x less than BEWO, while staying competitive or better on most metrics.

\begin{table}[t]
\centering
\small
\caption{Ablation study on the held-out MRSAudio set ($n=596$). Lower is
better for all metrics. PE: PE Core video features; I/T: ILD/ITD losses;
MD: modality dropout; $\dagger$: audio and video conditioning only (no
text). Bold marks the best score in each column.}
\label{tab:ablation}
\setlength{\tabcolsep}{3pt}
\begin{tabular*}{\linewidth}{@{\extracolsep{\fill}}lrrrr@{}}
\toprule
Model & FSAD & ITD & ILD & SCLAP\\
\midrule
PLACE & \textbf{5.14} & \textbf{0.11} & \textbf{1.98} & \textbf{0.04}\\
LoRA + PE & 6.30 & 0.11 & 2.17 & 0.18\\
LoRA only$^\dagger$ & 7.10 & 0.12 & 2.25 & 0.15\\
LoRA + PE + I & 7.22 & 0.12 & 2.36 & 0.21\\
LoRA + PE + I/T + MD & 8.96 & 0.13 & 2.58 & 0.23\\
LoRA + PE$^\dagger$ & 9.20 & 0.12 & 2.20 & 0.20\\
AudioX (base) & 44.67 & 0.22 & 3.00 & 0.29\\
\bottomrule
\end{tabular*}
\end{table}

\subsection{Ablation Studies}
\label{sec:ablat}
Each of the 596 held-out MRSAudio examples supplies a 10s video and accompanying caption;
video-only variants use a null (empty-string) text embedding. Training-only
modality dropout (MD) independently drops text, video, and audio with
probability 0.1 per stream, replacing text with its null embedding and other
streams with zeros, but never dropping text and video together. The spatial
branch always receives full video features. Table~\ref{tab:ablation} shows
that PE conditioning without interaural losses or MD gives the best
configuration without the adapter. Full PLACE adds the adapter trained
with both losses and MD and achieves the best performance across all four metrics.

\subsection{Subjective Listener Test}
\label{sec:test}
Our listener tests compare T2A outputs on five BEWO-1M SS-set and five
MRSAudio clips (GT, SpatialSonic, AudioX, PLACE) and V2A outputs on
five Gemini-generated videos and two held-out MRSAudio segments
(ViSAGe, AudioX, PLACE, and GT where available). Gemini videos test the
synthetic-video use case and have no GT. For MRSAudio examples, ViSAGe
is fine-tuned from its official checkpoint for two epochs. Each
participant rates five examples for semantic consistency, spatial
impression, spatial consistency, synchronization, and audio realism on
a 1--100 scale; T2A omits synchronization. Each T2A group has eleven
listeners and each V2A group ten. Table~\ref{tab:listener} separates
in-distribution MRSAudio from out-of-distribution Gemini and BEWO-1M
SS-set results; $n$ counts ratings per condition, with skipped examples
reducing the in-distribution V2A count. 

PLACE leads both V2A groups across all five dimensions and out-of-distribution T2A across all four, with its largest T2A margins in spatial impression and audio realism. For in-distribution T2A, PLACE trails both baselines overall: SpatialSonic leads on spatial impression and consistency, and AudioX on semantic consistency. Low GT ratings may reflect ambiguity between the caption and the audio in these examples. Together with the results in Secs.~\ref{sec:ood} and~\ref{sec:ablat}, PLACE’s strong V2A and T2A performance demonstrates the effectiveness of our data-efficient approach to binaural audio generation.
% We conduct subjective listening tests comparing binaural audio generated by T2A and V2A models. T2A uses five BEWO-1M SS-set test clips and five MRSAudio clips, comparing GT, SpatialSonic, AudioX, and PLACE. V2A uses five Gemini-generated videos and two held-out MRSAudio segments, comparing ViSAGe, AudioX, PLACE, and GT where available; Gemini videos have no GT. We use Gemini-generated videos in our evaluation because generating spatial audio for synthetic video is a common application of spatial audio models. ViSAGe is fine-tuned from its official checkpoint on MRSAudio for two epochs before generating MRSAudio examples. Each participant evaluates five examples, independently scoring semantic consistency, spatial impression, spatial consistency, synchronization, and audio realism from 1–100; synchronization is omitted for T2A. Each T2A group has eleven listeners and each V2A group ten. We group results by source into in-distribution (MRSAudio) and out-of-distribution (Gemini, BEWO-1M SS-set) conditions. In Table~\ref{tab:listener}, $n$ denotes ratings per condition; skipped examples reduce the in-distribution V2A count.

\begin{table}[t]
\centering
\small
\caption{Subjective listener evaluation. Ratings range from 1--100, higher
is better. SC: semantic consistency; SI: spatial impression; SP: spatial
consistency; Sy: synchronization (omitted for T2A); AR: audio realism;
Avg: overall average across the rated dimensions. GT: ground truth,
shown as a reference and excluded from bolding. Bold marks the best
model score in each column within each condition. \vspace{2pt}}
\label{tab:listener}
\setlength{\tabcolsep}{2pt}
\begin{tabular*}{\linewidth}{@{\extracolsep{\fill}}lrrrrrr@{}}
\toprule
Model & SC & SI & SP & Sy & AR & Avg\\
\midrule
\multicolumn{7}{@{}l}{\textit{T2A / In-distribution ($n=24$)}}\\
GT & 54.0 & 65.8 & 52.1 & -- & 69.6 & 60.4\\
PLACE & 38.3 & 38.6 & 31.3 & -- & 53.8 & 40.5\\
SpatialSonic & 44.2 & \textbf{48.2} & \textbf{42.5} & -- & 42.7 & \textbf{44.4}\\
AudioX & \textbf{49.2} & 33.0 & 27.8 & -- & \textbf{54.4} & 41.1\\
\midrule
\multicolumn{7}{@{}l}{\textit{T2A / Out-of-distribution ($n=25$)}}\\
GT & 73.6 & 68.3 & 64.4 & -- & 67.5 & 68.5\\
PLACE & \textbf{56.0} & \textbf{59.2} & \textbf{45.9} & -- & \textbf{61.0} & \textbf{55.5}\\
SpatialSonic & 46.6 & 49.5 & 44.2 & -- & 39.0 & 44.8\\
AudioX & 49.4 & 52.5 & 43.1 & -- & 52.4 & 49.3\\
\midrule
\midrule
\multicolumn{7}{@{}l}{\textit{V2A / In-distribution ($n=10$)}}\\
GT & 81.9 & 79.3 & 82.0 & 83.5 & 82.8 & 81.9\\
PLACE & \textbf{66.6} & \textbf{66.3} & \textbf{58.1} & \textbf{56.7} & \textbf{78.8} & \textbf{65.3}\\
ViSAGe & 44.1 & 46.4 & 39.6 & 48.7 & 45.7 & 44.9\\
AudioX & 29.0 & 33.5 & 28.2 & 35.8 & 31.7 & 31.6\\
\midrule
\multicolumn{7}{@{}l}{\textit{V2A / Gemini ($n=24$)}}\\
PLACE & \textbf{59.8} & \textbf{62.0} & \textbf{56.1} & \textbf{50.2} & \textbf{63.8} & \textbf{58.4}\\
AudioX & 58.1 & 47.8 & 45.2 & 43.4 & 63.2 & 51.5\\
ViSAGe & 13.5 & 14.4 & 14.7 & 15.1 & 24.6 & 16.5\\
\bottomrule
\end{tabular*}
\end{table}

\section{Conclusion}
PLACE extends AudioX with multimodal spatial conditioning and a latent
transformation learned through decoded-audio interaural losses. It supports
flexible binaural generation, improves on ViSAGe on most metrics,
and remains competitive with SpatialSonic on BEWO-1M SS-set test split despite training on a different
distribution. Reported listener ratings favor PLACE in overall averages.

\begin{center}\bf ACKNOWLEDGMENT\end{center}
We thank Gauri Jagatap for distributing the listener study within Dolby and all participants. We acknowledge the use of generative AI tools to assist with text and figure editing. 
The authors take full responsibility for all technical content, experiments, and results.

\clearpage % uncomment if necessary
\small % 9-point references.
\bibliographystyle{IEEEbib}
\bibliography{refs}

\begin{thebibliography}{10}

\bibitem{10.3389/frsip.2022.904866}
T.~Potter, Z.~Cvetković, and E.~De~Sena,
\newblock ``On the relative importance of visual and spatial audio rendering on {VR} immersion,''
\newblock {\em Frontiers in Signal Processing}, vol. Volume 2 - 2022, 2022.

\bibitem{garcia2026emorsion}
N.~Garcia, R.~Crocker, B.~M. {Del Sette}, F.~Smeraldi, C.~Saitis, G.~Fazekas, and J.~Reiss,
\newblock ``{EMORSION}: Examining the impact of audio parameters on emotional responses and immersion in film,''
\newblock in {\em Proc. AES Europe Convention}, 2026,
\newblock arXiv:2606.18266.

\bibitem{kim2025visage}
J.~Kim, H.~Yun, and G.~Kim,
\newblock ``{ViSAGe}: Video-to-spatial audio generation,''
\newblock in {\em ICLR}, 2025.

\bibitem{zhao2026foleyspace}
L.~Zhao, R.~Chen, C.~Zhang, X.-L. Zhang, and X.~Li,
\newblock ``{FoleySpace}: Vision-aligned binaural spatial audio generation,''
\newblock {\em IEEE Transactions on Multimedia}, pp. 1--11, 2026.

\bibitem{sun2024both}
P.~Sun, S.~Cheng, X.~Li, Z.~Ye, H.~Liu, H.~Zhang, W.~Xue, and Y.~Guo,
\newblock ``Both ears wide open: Towards language-driven spatial audio generation,''
\newblock in {\em International Conference on Learning Representations}, 2025, vol. 2025, pp. 42606--42649.

\bibitem{zhang2025visaudioendtoendvideodrivenbinaural}
M.~Zhang, Q.~Chen, T.~Wu, Z.~Liu, and D.~Lin,
\newblock ``{ViSAudio}: End-to-end video-driven binaural spatial audio generation,'' arXiv preprint arXiv:2512.03036, 2025.

\bibitem{lei2026swansphere}
K.~Lei, Y.~Zhang, C.~Pan, X.~Pu, W.~Guo, R.~Li, and Z.~Zhao,
\newblock ``Towards streaming synchronized spatial audio generation via autoregressive diffusion transformer,''
\newblock in {\em Proceedings of the 43rd International Conference on Machine Learning}. 2026, Proceedings of Machine Learning Research, PMLR,
\newblock arXiv:2605.30940.

\bibitem{gao2019visualsound}
R.~Gao and K.~Grauman,
\newblock ``{2.5D} {Visual Sound},''
\newblock in {\em CVPR}, 2019.

\bibitem{tian2025audiox}
Z.~Tian, Z.~Liu, Y.~Jin, R.~Yuan, L.~Xue, X.~Tan, Q.~Chen, W.~Xue, and Y.~Guo,
\newblock ``{AudioX}: A unified framework for anything-to-audio generation,''
\newblock in {\em International Conference on Learning Representations}, C.~Vondrick, B.~Hariharan, C.~Raffel, L.~Pinto, D.~Yang, and A.~Faust, Eds., 2026, vol. 2026, pp. 110625--110651.

\bibitem{evans2025stable}
Z.~Evans, J.~D. Parker, C.~Carr, Z.~Zukowski, J.~Taylor, and J.~Pons,
\newblock ``{Stable Audio Open},''
\newblock in {\em ICASSP 2025-2025 IEEE International Conference on Acoustics, Speech and Signal Processing (ICASSP)}. IEEE, 2025, pp. 1--5.

\bibitem{synchformer2024iashin}
V.~Iashin, W.~Xie, E.~Rahtu, and A.~Zisserman,
\newblock ``Synchformer: Efficient synchronization from sparse cues,''
\newblock in {\em ICASSP 2024-2024 IEEE International Conference on Acoustics, Speech and Signal Processing (ICASSP)}. IEEE, 2024.

\bibitem{radford2021learning}
A.~Radford, J.~W. Kim, C.~Hallacy, A.~Ramesh, G.~Goh, S.~Agarwal, G.~Sastry, A.~Askell, P.~Mishkin, J.~Clark, et~al.,
\newblock ``Learning transferable visual models from natural language supervision,''
\newblock in {\em International conference on machine learning}. PmLR, 2021, pp. 8748--8763.

\bibitem{T5}
C.~Raffel, N.~Shazeer, A.~Roberts, K.~Lee, S.~Narang, M.~Matena, Y.~Zhou, W.~Li, and P.~J. Liu,
\newblock ``Exploring the limits of transfer learning with a unified text-to-text transformer,''
\newblock {\em J. Mach. Learn. Res.}, vol. 21, no. 1, Jan. 2020.

\bibitem{bolya2025PerceptionEncoder}
D.~Bolya, P.-Y. Huang, P.~Sun, J.~H. Cho, A.~Madotto, C.~Wei, T.~Ma, J.~Zhi, J.~Rajasegaran, H.~A. Rasheed, J.~Wang, M.~Monteiro, H.~Xu, S.~Dong, N.~Ravi, S.-W. Li, P.~Dollar, and C.~Feichtenhofer,
\newblock ``Perception encoder: The best visual embeddings are not at the output of the network,''
\newblock in {\em The Thirty-ninth Annual Conference on Neural Information Processing Systems}, 2025.

\bibitem{wang2024internvideo2}
Y.~Wang, K.~Li, X.~Li, J.~Yu, Y.~He, G.~Chen, B.~Pei, R.~Zheng, Z.~Wang, Y.~Shi, et~al.,
\newblock ``{InternVideo2}: Scaling foundation models for multimodal video understanding,''
\newblock in {\em European conference on computer vision}. Springer, 2024, pp. 396--416.

\bibitem{laionclap2023}
Y.~Wu, K.~Chen, T.~Zhang, Y.~Hui, T.~Berg-Kirkpatrick, and S.~Dubnov,
\newblock ``Large-scale contrastive language-audio pretraining with feature fusion and keyword-to-caption augmentation,''
\newblock in {\em IEEE International Conference on Acoustics, Speech and Signal Processing, ICASSP}, 2023.

\bibitem{losses}
C.~Hernandez-Olivan, M.~Delcroix, T.~Ochiai, N.~Tawara, T.~Nakatani, and S.~Araki,
\newblock ``Interaural time difference loss for binaural target sound extraction,''
\newblock in {\em 2024 18th International Workshop on Acoustic Signal Enhancement (IWAENC)}, 2024, pp. 210--214.

\bibitem{ick2024spatiallyaware}
C.~A. Ick, G.~Wichern, Y.~Masuyama, F.~Germain, and J.~{Le Roux},
\newblock ``Spatially-aware losses for enhanced neural acoustic fields,''
\newblock in {\em Audio Imagination: NeurIPS 2024 Workshop AI-Driven Speech, Music, and Sound Generation}, 2024.

\bibitem{knapp1976generalized}
C.~Knapp and G.~Carter,
\newblock ``The generalized correlation method for estimation of time delay,''
\newblock {\em IEEE transactions on acoustics, speech, and signal processing}, vol. 24, no. 4, pp. 320--327, 1976.

\bibitem{hu2022lora}
E.~J. Hu, Y.~Shen, P.~Wallis, Z.~Allen-Zhu, Y.~Li, S.~Wang, L.~Wang, and W.~Chen,
\newblock ``Lo{RA}: Low-rank adaptation of large language models,''
\newblock in {\em International Conference on Learning Representations}, 2022.

\bibitem{guo2025mrsaudio}
W.~Guo, C.~Pan, Z.~Zhu, X.~Hu, Y.~Zhang, L.~Tang, R.~Yang, H.~Wang, Z.~Zhang, Y.~Wang, Y.~Chen, H.~Xu, K.~Xu, P.~Fan, Z.~Chen, Y.~Yu, Q.~Huang, F.~Wu, and Z.~Zhao,
\newblock ``{MRSA}udio: A large-scale multimodal recorded spatial audio dataset with refined annotations,''
\newblock in {\em The Thirty-ninth Annual Conference on Neural Information Processing Systems Datasets and Benchmarks Track}, 2026.

\bibitem{hold2025spaudiopy}
C.~Hold,
\newblock ``{spaudiopy}: Spatial audio python package,'' Software, version 0.2.0, 2025,
\newblock [Online]. Available: \url{https://pypi.org/project/spaudiopy/}. Accessed: Sep. 23, 2026.

\bibitem{arihrtf}
{Acoustics Research Institute, Austrian Academy of Sciences},
\newblock ``{ARI} head-related transfer function database,'' Online database,
\newblock [Online]. Available: \url{https://www.oeaw.ac.at/en/ari/outreach/software/hrtf-database}. Accessed: Sep. 23, 2026.

\bibitem{aes69}
{Audio Engineering Society},
\newblock ``{AES69-2015}: Aes standard for file exchange---spatial acoustic data file format,''
\newblock Tech. {R}ep., Audio Engineering Society, 2015.

\bibitem{cheng2025taming}
H.~K. Cheng, M.~Ishii, A.~Hayakawa, T.~Shibuya, A.~Schwing, and Y.~Mitsufuji,
\newblock ``{MMAudio}: Taming multimodal joint training for high-quality video-to-audio synthesis,''
\newblock in {\em CVPR}, 2025.

\bibitem{chen2022sound}
Z.~Chen, D.~F. Fouhey, and A.~Owens,
\newblock ``Sound localization by self-supervised time delay estimation,''
\newblock in {\em European Conference on Computer Vision}. Springer, 2022, pp. 489--508.

\bibitem{11460767}
K.~Seki, Y.~Okamoto, K.~Yamaoka, Y.~Saito, S.~Takamichi, and H.~Saruwatari,
\newblock ``{Spatial-CLAP}: Learning spatially-aware audio–text embeddings for multi-source conditions,''
\newblock in {\em ICASSP 2026 - 2026 IEEE International Conference on Acoustics, Speech and Signal Processing (ICASSP)}, 2026, pp. 14742--14746.

\end{thebibliography}
\end{document}